\AtBeginDocument{%
  \paperwidth=\dimexpr
    1in + \oddsidemargin
    + \textwidth
    + 1in + \oddsidemargin
  \relax
  \paperheight=\dimexpr
    1in + \topmargin
    + \headheight + \headsep
    + \textheight
    + 1in + \topmargin
  \relax
  \usepackage[pass]{geometry}\relax
}

\RequirePackage{fix-cm}
\documentclass[smallextended]{svjour3}       
\smartqed  

\usepackage{hyperref} 

\usepackage{ragged2e}
\usepackage[figuresleft]{rotating}
\usepackage{booktabs}
\usepackage{url}
\usepackage{balance}
\usepackage{breqn}
\usepackage{array}
\usepackage{colortbl,booktabs}
\usepackage{subcaption}
\usepackage{titlesec}
\usepackage{algorithmic}
\usepackage{algorithm}
\usepackage{dblfloatfix}
\usepackage{mathtools,amssymb}
\usepackage{tabularx}
\usepackage{pgfplots} 
\usetikzlibrary{patterns}
\usepackage{graphicx}
\usepackage{multirow}

\newcommand{\etal}{\textit{et al. }}

\begin{document}

\titlerunning{Can we Quantify Trust? Towards a Trust-based Resilient SIoT Network}

\title{Can we Quantify Trust? Towards a Trust-based Resilient SIoT Network}

\author{{Subhash Sagar\thanks{Subhash Sagar's research is funded via the Macquarie University Research Excellence Award (Allocation No. 2019050). Adnan Mahmood's research has been supported under the auspices of the Macquarie University's COVID Recovery Research Fellowship Grant 180420387. Quan Z. Sheng’s work has been partially supported via the Australian Research Council Future Fellowship Grant FT140101247 and Discovery Project Grants DP200102298.}} \and Adnan Mahmood \and Quan Z. Sheng \and Munazza Zaib \and Farhan Sufyan}

\authorrunning{Subhash Sagar et al.}

\institute{{Subhash Sagar, Adnan Mahmood, Quan Z. Sheng, Munazza Zaib \at
              School of Computing, Macquarie University, NSW 2109, Australia. 
          }\\
          {Farhan Sufyan \at
              School of Computing Science and Engineering, Galgotias University, India. 
          } 
}


\maketitle

\begin{abstract}

The emerging yet promising paradigm of the Social Internet of Things (SIoT) integrates the notion of the Internet of Things with human social networks. In SIoT, objects, i.e., \emph{things}, have the capability to socialize with the other objects in the SIoT network and can establish their social network autonomously by modeling human behaviour. The notion of trust is imperative in realizing these characteristics of socialization in order to assess the reliability of autonomous collaboration. The perception of trust is evolving in the era of SIoT as an extension to traditional security triads in an attempt to offer secure and reliable services, and is considered as an imperative aspect of any SIoT system for minimizing the probable risk of autonomous decision-making. This research investigates the idea of trust quantification by employing trust measurement in terms of direct trust, indirect trust as a recommendation, and the degree of SIoT relationships in terms of social similarities (community-of-interest, friendship, and co-work relationships). A weighted sum approach is subsequently employed to synthesize all the trust features in order to ascertain a single trust score. The experimental evaluation demonstrates the effectiveness of the proposed model in segregating trustworthy and untrustworthy objects and via identifying the dynamic behaviour (i.e., trust-related attacks) of the SIoT objects.


\end{abstract}

\section{Introduction}\label{sec1}

The notion of the Internet of Things (IoT) refers to the billions of smart objects (e.g., gadgets, machines, and associated software) equipped with sensors and actuators, connected to the internet \cite{ashton2009internet}\cite{ATZORI20102787}. This evolution of connected smart objects has led to a number of promising real-world applications, having direct inference on our daily lives, and such applications include smart cities, smart healthcare, smart homes, etc \cite{9319033}. It is anticipated by Statista\footnote{https://www.statista.com/statistics/1101442/iot-number-of-connected-devices-worldwide/} that by 2025, there will be around more than 30 billion smart objects, and as a result, scalability and navigability are some of the significant challenges to the adoption of the IoT ecosystem. 

The paradigm of the Social Internet of Things (SIoT) is a promising solution to address such challenges. The notion of SIoT has augmented the idea of IoT by incorporating the concept of social networking in smart objects, wherein each object can establish social relationships with other objects autonomously based on the rules set out by their respective owners \cite{ATZORI20123594}. Some of the fundamental SIoT relationships can fall into the category of ownership object relationships, social object relationships, parental object relationships, co-location objects relationships, and co-work object relationships. The socialization of objects (via SIoT relationships) has paved the way for the next generation of IoT with an ability to accommodate trillions of smart objects (i.e., service requestors and providers), and has led to numerous benefits, including but not limited to assurance of effective service discovery and network navigability, network scalability similar to human beings, establishing trustworthy relationships among objects, and utilizing of social network architecture for SIoT system. Nevertheless, maintaining trustworthy relationships and providing seamless connectivity to a multitude of heterogeneous objects is always fraught with risk owing to the security and trust of these objects \cite{7576667} \cite{HOSSEINISHIRVANI2023100640}. Since SIoT services are expected to make the decision autonomously without any human intervention, it is imperative for the service requester (i.e., trustor) to determine the trustworthiness of the objects before relying on the information provided by a service provider (i.e., trustee). This kind of trust assessment is essential since there are malevolent objects inside the network that are primarily motivated by the intent to jeopardize the network resources for harmful goals, for instance, the dissemination of malware or false information.

Given the aforementioned insights, the motive for establishing trustworthiness management for SIoT is indisputable. Over the past few years, a number of studies have been proposed in an effort to address the challenges of trustworthiness management in a variety of disciplines including but not limited to mobile and vehicular ad hoc networks \cite{10.1145/3594637}, peer-to-peer networks \cite{7407618}, online social networks (for the identification of malicious users and sometimes fake stories) \cite{8737469}, and e-commerce (wherein the credibility of a service provider (i.e., retailer) is shared by users by the means of transactions) \cite{6595455}. The notion of trust in SIoT is characterized as the expectation of a trustor on a trustee to accomplish a well-defined objective in a particular domain within a specific time period. Trust assessment can be a value or a probability and is not the property of either trustor or a trustee, in fact, it is a correlation between the two within a particular environment. Moreover, trust assessment requires a substantial number of parameters owing to its complex dynamics that varies with environments and their respective contexts.  

\begin{figure}[!t]
    \centering
    \includegraphics[width=0.8\linewidth]{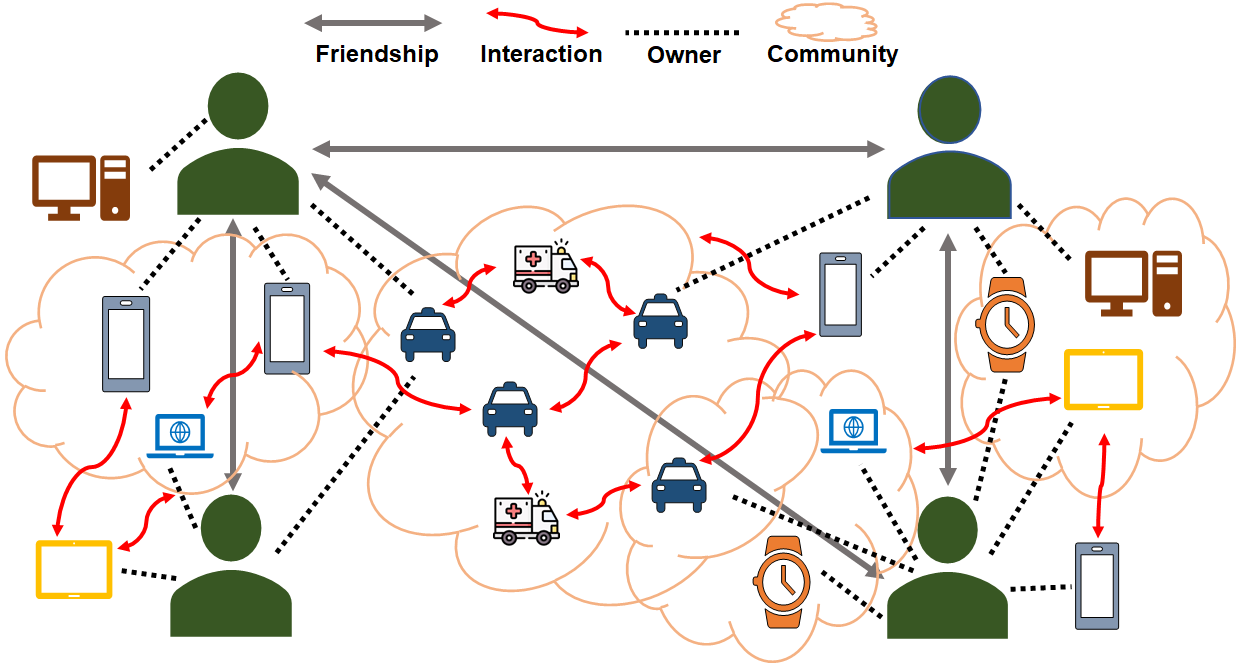}
    \caption{A high-level view of similarity-based SIoT model.}
    \label{fig:Siot_model_ch2}
\end{figure}

Accordingly, a SIoT-specific similarity-based trust quantification model is therefore proposed to measure the trust score of a SIoT object in this research. An illustration of the SIoT network encompassing a number of similarities is depicted in Figure \ref{fig:Siot_model_ch2} \cite{9322540}, wherein objects interact with one another in a highly decentralized manner. In addition, objects are aligned to function in certain communities or at the workplace and have owners who keep a list of friends and communities-of-interest to symbolize social interactions in terms of social similarities. Direct and indirect (i.e., recommendations) interactions are necessary for building the trust of an object. Additionally, the perception or trust of one object for another object in the SIoT network is updated based on their interactions at any particular time instance. Furthermore, each object is accountable for independently carrying out the trust quantification process and defining its direct trust perception for other objects upon encounter by utilizing its owner's friendships, communities-of-interest, co-work relationships, and interactions amongst them. Finally, the main contributions of this study are as follows:

\begin{itemize}
  \item[--] A SIoT-specific similarity-based trust quantification model has been envisaged by employing direct perception (i.e., direct trust), indirect perception (i.e., recommendation/indirect trust), and the social characteristics in terms of social similarity of trustor-trustee pair in order to embark the misbehaving objects in the network whose status changes with varying interactions; 
  \item[--] A weighted sum scheme has been envisaged to aggregate the trust features for a unified trust score, wherein a combination of weight schemes are employed in order to efficiently aggregate the employed trust features and to analyze the suitable amalgamation of weights; and
  \item[--] Finally, the experimental evaluation of the proposed model has been conducted in a simulation environment with a different number of interactions in order to monitor the trust score of the benevolent and malevolent objects. Furthermore, we have also analyzed the trust-based dynamically changing behaviour of objects throughout the interactions with varying weights schemes.
\end{itemize}
\vspace{-0.5em}

The remainder of the paper is organized as follows. Section \ref{sec:back_sota} presents the background of trust and an overview of the existing state-of-the-art trust computational model for SIoT. Section \ref{sec:proposed_model} provides the detail of the employed trust quantification model for SIoT. Section \ref{sec:sim_res_ch2} reports the simulation setup and experimental results for the performance evaluation of the proposed model. Finally, Section \ref{sec:summ_ch2} gives the concluding remarks.
\vspace{-0.75em}

\section{Background and State-of-the-Art}
\label{sec:back_sota}
\subsection{Background}

The idea of trustworthiness management in SIoT is evolving rapidly, and it is, therefore, indispensable to know the ideal trust parameter for any SIoT system. This section delineates the notion of trust and its perspective in the SIoT and the current state-of-the-art in trustworthiness management for the SIoT. 

\subsubsection{Trust Concept}
The notion of trust is a fundamental aspect of human society and with the advancement in science (e.g., in terms of software and hardware), the concept of trust has been utilized in a number of disciplines (i.e., sociology, psychology, economics and computer science) \cite{926617} \cite{9264256}. The concept of trust differs across disciplines and the fundamental definition of trust is "the confidence of a trustor in a trustee," and perceptions of trust depend on a variety of facets, including but not limited to, temporal factors, environmental factors, and human propensities \cite{sagar2022understanding}. In computer science, trust is considered as network and information security, and a system is believed to be trustworthy if it is secure and can categorize the individual accessing a particular system in order to guarantee the integrity and privacy of the information. The early variant of trust in computer science is characterized as a UNIX program free from Trojan horses \cite{358210}.  

\subsubsection{Trust in SIoT}

The foundation of the SIoT paradigm is focused on social interactions and is more inclined towards social science and trust is a crucial component of human social interactions. According to a widely accepted definition in social science, trust is defined as ``confidence" or ``self-assurance". As a result, trust in the context of the SIoT is often understood to be the confidence of a trustor in a trustee to achieve a goal within a certain context and within a specific time frame.

The measure of trust as confidence (also known as trust esteem) can be a probability or a value in the context of the SIoT. An object is also referred to as a trustor or trustee and can be a person, a machine, or an application. Furthermore, it's crucial to comprehend that a trust is a relationship between the trustor and the trustee rather than being either of their possessions \cite{app9010166}. The overarching goal of the trust is understood as a trustee's action, or it might be the information that the trustee provides based on the expectations of the trustor and the trustee's personal characteristics \cite{article}\cite{9917502}. The key components in quantifying an SIoT object's trust score are knowledge extraction (using social trust features or Quality-of-Service features), trust aggregation (using traditional weighted sum, fuzzy logic, machine learning, etc.), and finally, trust decision, which determines whether an object is trustworthy or not \cite{9762320}\cite{riz2022}\cite{10054446}.\vspace{-1.5em}

\subsection{State-of-the-Art}

Recent years suggest the extensive utilization of the trust concept as an essential aspect of any SIoT system. Accordingly, a context-aware socio-cognitive-based trust model for service delegation in service-oriented SIoT is proposed by Wei \etal \cite{9211717}, wherein two characteristics \textit{competence quantification} and \textit{willingness quantification} form the basis of the model. Furthermore, The degree of importance and the degree of social connections (DoSR) are used to quantify competence, and the degree of contribution (DoC) and the DoSR are also incorporated in the measurement of willingness. The DoC guarantees the service provider's willingness, the DoI measures the competency of service providers in terms of processing power, storage, and communication capabilities, and the DoSR is used as the weighing criteria for both competence and willingness. In essence, the weighted sum approach is employed to aggregate the two trust parameters in order to provide the final trust score. Similarly, Pourmohseni \etal in \cite{POURMOHSENI2022758} delineated a trust model for SIoT by employing a variety of trust parameters, (i.e., QoS, social and context-based). Nevertheless, a new perspective for trust quantification is discussed which integrates the neutrosophic numbers with the trust-related data in order to deal with the uncertainty and inconsistency in trust-related data before quantifying the selected trust parameters.  Finally, the weighted-sum aggregation is utilized to get the single trust score. 

Furthermore, trustworthiness management systems are utilized for a number of IoT applications. For instance, in \cite{8662641}, a trust evaluation mechanism is proposed for recruiting mobile nodes for crowdsourcing, wherein two trust parameters, namely experience and reputation are used and aggregated to compute the trust score of a node. Similarly, a recommendation-based trust model for vehicle-to-everything (V2X) communication is provided in \cite{8887207}. The suggestion from nearby nodes (i.e., automobiles and/or roadside units) determines how the weights are updated when combining direct trust and recommendation to determine the trust score of each vehicle. In addition, Mohammadi \etal \cite{MOHAMMADI2021107479} delineated a trust-based friends selection algorithm using an exhaustive search. The SIoT relationships (e.g., parental object relationships, social relationships, etc.) are employed to select trustworthy friends, wherein the data profiling, distance, and interactions are considered in terms of probability distribution to ascertain the degree of SIoT relationships. Finally, two types of trust scores (static and dynamic) are considered to quantify the trust score in order to eliminate the untrustworthy SIoT objects. 

A machine learning-based trust framework based on a node's social profile has been developed by Jayasinghe \etal \cite{8364607}, whereby various social characteristics are accumulated by utilising machine learning-based techniques to obtain the direct trust metric of any node in an IoT network. Similarly, a deep learning-based trust resilient model is proposed by Magdich \etal \cite{MAGDICH20} in order to not only mitigate the trust-related attacks in terms of service provider’s behaviour but also detect poor service providers. Furthermore, Xia \etal \cite{8737491} delineate a trustworthiness inference framework by employing two trust measures, \textit{similarity trust} and \textit{familiarity trust}. Subsequently, a fuzzy logic-based aggregation technique is proposed to synthesize both trust metrics in order to get a single trust score. Most recently, an artificial neural network-based trustworthy object classification model for SIoT (referred to as ``\textit{Trust-SIoT}'') is delineated that considers a number of trust features, i.e., direct trust, indirect trust as a recommendation, the credibility of the recommending objects in terms of their reliability and benevolence, and the social similarity, to classify the SIoT objects as trustworthy, untrustworthy or neutral \cite{10054446}. As of late, a number of studies staged the idea to employ blockchain-based trust models \cite{10.1007/s11227-021-04231-3}\cite{9120287}\cite{Alam2022}, e.g., a lightweight blockchain-based trust evaluation mechanism is introduced \cite{10.1007/s11227-021-04231-3}, wherein the SIoT relationships among the objects are considered in the form of a social network. Moreover, an Ethereum platform is utilized to realize the validity of the model in detecting the untrustworthy SIoT object performing trust-related attacks. Nevertheless, the model still needs the fundamental trust metrics, i.e., direct trust, indirect trust, and social relationships to compute the trust score of SIoT devices. It is evident that the recent advancement in technology has the potential to be employed in the trustworthiness management system. however, integrating the concept of context-awareness, the dynamic nature of SIoT application, and the computational latency are some of the challenges that need further exploration. 
\vspace{-2em}
\section{Trust Quantification Model}
\label{sec:proposed_model}

\begin{figure}[!t]
    \centering
    \includegraphics[width=0.75\linewidth]{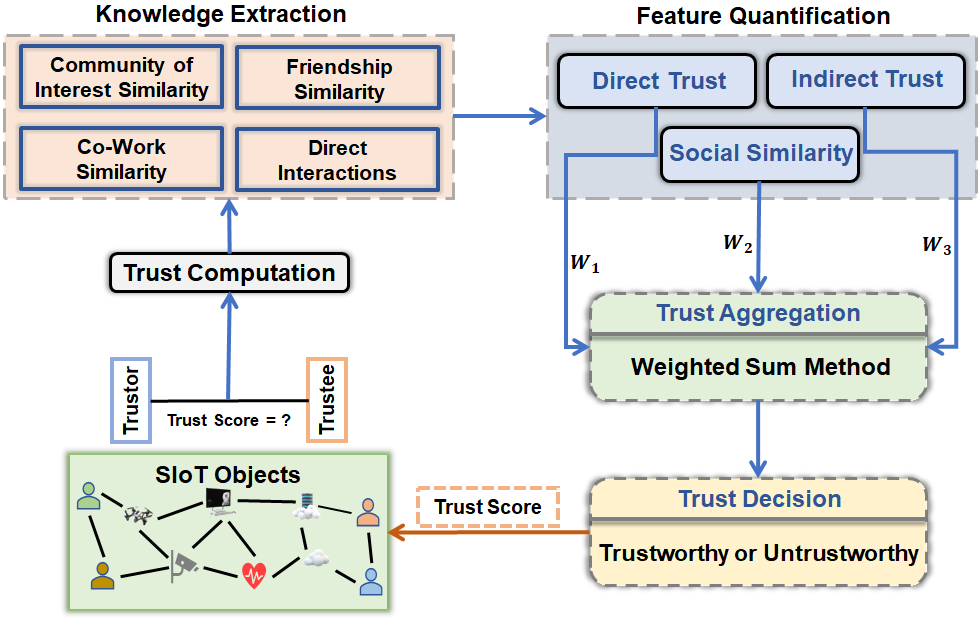}
    \caption{A schematic diagram of the proposed trust computational model.}
    \label{fig:trust_quantify}
    \vspace{-1.5em}
\end{figure}

As depicted in Figure~\ref{fig:trust_quantify}, the envisaged trust quantification model considers all the characteristics of trustworthiness management, including but not limited to, knowledge extraction, quantification of trust features from the knowledge in terms of the direct trust (i.e., direct observation), indirect trust (i.e., recommendations), and the degree of social similarity, followed by trust aggregation, and finally, the trust decision. Knowledge extraction from the SIoT network is the first step in trustworthiness management, the proposed trust quantification model extracts the SIoT relationships in terms of social metrics (i.e., Community-of-interest, Friendship, and Co-work Similarity) and the information of direct interactions in terms of positive and negative interactions. Subsequently, the feature extraction step quantifies the extracted knowledge in terms of direct trust, indirect trust, and the degree of social similarity. A weighted sum approach is then employed to aggregate all the extracted trust features in order to obtain the single trust score. Finally, it is a trust decision step that is responsible for the classification of the SIoT object into trustworthy or untrustworthy groups via trust-threshold value ($\theta$). The final trust score of an object (i.e., trustor) $i$ towards another object (i.e., trustee) $j$ in the SIoT network is denoted by $Trust_{FT}^t (i,j)$ at time $t=[0,t]$. The final trust score encompasses three trust observations – (i) Direct Trust ($Trust_{DT}^t (i,j)$), (ii) Indirect Trust or Recommendations ($Trust_{RT}^t (i,j)$), and (iii) Social Similarity ($Trust_{SS}^t (i,j)$). The range of final trust score varies between $[0,1]$, wherein the score closer to $0$ classifies the object as untrustworthy and the score closer to $1$ classifies an object as trustworthy.

\vspace{-1.5em}

\subsection{Direct Trust ($Trust_{DT}^t$)}

Direct trust represents the direct observation of a trustor $i$ towards a trustee $j$. The proposed model quantifies the direct observations by employing both successful (positive) interactions and unsuccessful (negative) interactions between a trustor-trustee pair. We have considered the Bayesian inference with beta probability density function to quantify the direct trust \cite{Ismail2002TheBR}. The direct trust of a trustor towards a trustee is defined as:

\begin{equation}
    Trust_{DT}^t (i,j) = \frac{\mathcal{P}^t (i,j) + 1}{\mathcal{P}^t (i,j) + \mathcal{N}^t (i,j) + 2}
    \label{eq:dt_ch2}
\end{equation}
wherein, $\mathcal{P}$ and $\mathcal{N}$ represent positive and negative interactions respectively at any given time $t$. These positive and negative interactions represent the feedback provided by the trustor and are considered one of the key characteristics of the trust quantification process. In general, it is presumed that a trustor can perfectly rate the trustee (i.e., received service) after the service is fully realized.  
\vspace{-1em}

\subsection{Indirect Trust -- Recommendation ($Trust_{R}^t$)}
In contrast to direct trust, the idea of indirect trust is to provide the recommendation as a trust about a trustee to a trustor in the absence of direct observation vis-à-vis the trustor-trustee pair. Furthermore, the recommendation is an indispensable indicator if the trustor needs the recommendations from the neighbouring objects in the SIoT network and is effective in order to accurately quantify the trust score of a trustee. The proposed model employs the mean of the direct trust of neighbouring friends $k$ towards the trustee $j$ in order to ascertain the recommendations as a trust $Trust_{R}^t$ at time $t$ as is computed as:

\begin{equation}
        Trust_{R}^t (k,j) = \frac{1}{N} \sum_{N=1}^{m} Trust_{DT}^t(k_N,j)
    \label{eq:rt_ch2}
\end{equation}

\noindent wherein, $N$ signifies the total number of neighbouring objects. Furthermore, the proposed model has taken into consideration the recommendations from the neighbouring friends of the trustor having a direct trust score of above the threshold ($\theta$) to address the issues of a variety of trust-related attacks (i.e., bad-mouthing attacks and ballot-stuffing attacks) during the amalgamation of recommendations as a trust with the final trust score.  
\vspace{-1em}

\subsection{Social Similarity ($Trust_{SS}^t$)}
The social similarity feature $Trust_{SS}^t$  is employed to ascertain the social aspects of a trustee towards a trustor at any time t. In essence, the social aspects of a trustee could be assessed by utilizing a number of measures, the proposed model exploits three fundamental similarities metrics to assess a trustee and are described as follows:

\subsubsection{Community-of-Interest Similarity ($Sim_{CoI}^t$)}
This trust feature determines the similarity in interests between a trustor $i$ and a trustee $j$, by determining the degree of community-based similarity. It is achieved by comparing the common interests, such as memberships in similar online social networking and e-commerce groups, between the trustor and trustee. The community-based similarity is calculated by taking the ratio of common communities in which both trustor and trustee are active to the total number of communities in which both parties are involved. This community-of-interest similarity $Sim CoIt(i,j)$ at time $t=[0,t]$ is ascertained as follows:

\begin{equation}
    Sim_{CoI}^t (i,j) = \frac{|C_i \cap C_j|}{|C_i \cup C_j|}
    \label{eq:coi_ch2}
\end{equation}

\noindent where, $C_i$ and $C_j$ represent the set of communities of a trustor and a trustee respectively, and $|.|$ shows the cardinality of a set, i.e., count of the communities. 

\subsubsection{Friendship Similarity ($Sim_{FS}^t$)}

The friendship similarity signifies the importance of an object in terms of its social relationships with other neighbouring objects (i.e., friends). The primary intent of friendship similarity is to assess the significance of an object to prohibit malicious objects from establishing forged relationships. In essence, the friendship similarity ($Sim_{CoI}^t (i,j)$) as the ratio of common friends between a trustor-trustee pair to the total number of friends a trustor $i$ and a trustee $j$ at time $t=[0,t]$ is ascertained as follows:

\begin{equation}
    Sim_{FS}^t (i,j) = \frac{|F_i \cap F_j|}{|F_i \cup F_j|}
    \label{eq:fs_ch2}
\end{equation}

\noindent where, $F_i$ and $F_j$ represent the set of friends of a trustor and a trustee respectively.

\subsubsection{Co-work Similarity ($Sim_{CW}^t$)}

The co-work similarity feature of an object is measured when the functionality of two or more objects is integrated to achieve a shared purpose by collaborating in a common SIoT application. In this case, the co-work relationships between the SIoT object are prioritized over their physical location. The cosine similarity between the multicast interactions of the trustor $i$ and the trustee $j$ is used to measure the degree of co-work relationships. In essence, the co-work similarity $Sim_{CW}^t (i,j)$ is considered as the ratio of common multicast interactions vis-à-vis trustor-trustee pair to the total number of multicast interactions and is computed as: 

\begin{equation}
    Sim_{CW}^t (i,j) = \frac{|M_i \cap M_j|}{\sqrt{|M_i|.|M_j|}}
    \label{eq:cw_ch2}
\end{equation}

Finally, the social similarity ($Trust_{SS}^t (i,j)$) as the trust metric vis-à-vis trustor-trustee pair is thus computed as follows:

\begin{equation}
        Trust_{SS}^t (i,j) = \frac{1}{n}\sum_{f=1}^{n} Sim_{\mathcal{X}}^t(i,j)
    \label{eq:total_ch2}
\end{equation}

\noindent wherein, $\mathcal{X}$ represents the degree of social similarity (i.e., $CoI$, $FS$, and $CW$) and $n$ signifies the count of integrated similarity measures.
\vspace{-0.5em}

\subsection{Final Trust Score}

Conclusively, a weighted sum method approach is employed to combine all the trust features in order to ascertain a single trust value and is depicted as:

\begin{equation}
    Trust_{FT}^t (i,j) = w_1*Trust_{DT}^t (i,j) + w_2*Trust_{SS}^t (i,j) +w_3*Trust_{R}^t (i,j)
    \label{eq:final_t_ch2}
\end{equation}

\noindent here, $w$ signifies the weighting factors and in the proposed model, we have identified and compared a combination of weights for the final trust score. In particular, three different weight schemes (Table \ref{tab:weights_ch2}) are utilized in order to ascertain the final trust score. In essence, the weight parameters employ the importance of each trust feature in obtaining the final trust score. The proposed model employs three variants of the weight schemes, Weight Scheme-1 ($WS-1$) \cite{9211717}, Weight Scheme-2 ($WS-2$) \cite{Nitti2014}, and equal weights ($Mean$) \cite{6940301}, i.e., mean as the baseline approach of selecting weights. 

\renewcommand\arraystretch{1.5}
\begin{table}[t]
    \centering
\begin{tabular}{ c |c c c} 
\hline \hline
\rowcolor{gray!30}  & \multicolumn{3}{c}{Weights} \\ \cline{2-4}
\rowcolor{gray!30} \multirow{-2}{*}{Features} & $WS-1$ & $WS-2$ & $Mean$ \\
\hline \hline
            $Trust_{DT}^t(i,j)$ & 0.5 & 0.4 & 0.33 \\ \hline
            $Trust_{SS}^t(i,j)$ & 0.3 & 0.3 & 0.33 \\ \hline
            $Trust_{R}^t(i,j)$ & 0.2 & 0.3 & 0.33 \\
\hline \hline
\end{tabular}
\caption{Weight Schemes (WS)}
\label{tab:weights_ch2}
\vspace{-0.5em}
\end{table}

The final trust computation needs to consider the following possible scenarios to quantify the trust of an object:

\begin{itemize}
     \item[1.] All the trust features, $Trust_{DT}^t(i,j)$, $Trust_{SS}^t(i,j)$, and $Trust_{R}^t(i,j)$ are available. 
    \item[2.] There is no $Trust_{SS}^t(i,j)$ but $Trust_{DT}^t(i,j)$ and $Trust_{R}^t(i,j)$ are available. 
    \item[3.] There is no $Trust_{R}^t(i,j)$ but $Trust_{DT}^t(i,j)$ and $Trust_{SS}^t(i,j)$ are available. 
    \item[4.] There is no $Trust_{DT}^t(i,j)$ but $Trust_{SS}^t(i,j)$ and $Trust_{R}^t(i,j)$ are available.
    \item[5.] Only $Trust_{DT}^t(i,j)$ is available. 
    \item[6.] Only $Trust_{R}^t(i,j)$ is available.
\end{itemize}

Furthermore, each scenario considers a specific combination of trust features, therefore the weights assignments of each scenario are different and can be seen in Table \ref{tab:final_trust_weights_ch2}.
\vspace{-1em}

\begin{table}[ht]
\begin{center}
\begin{tabular}{ c|c|c|c}
\hline \hline
\rowcolor{gray!30}
Scenario & $Trust_{DT}^t(i,j)$ & $Trust_{SS}^t(i,j)$ & $Trust_{R}^t(i,j)$ \\ \hline \hline
1 & $w_1$ & $w_2$ & $w_3$ \\ \hline
2 & $w_1+w_2$ & 0 & $w_3$ \\ \hline
3 & $w_1+w_3$ & $w_2$ & 0 \\ \hline
4 & 0 & $w_2$ & $w_3+w_1$ \\ \hline
5 & $w_1+w_2+w_3$ & 0 & 0 \\ \hline
6 & 0 & 0 & $w_1+w_2+w_3$ \\ \hline \hline
\end{tabular}
\caption{Weights for each scenario to compute the final trust score}
\label{tab:final_trust_weights_ch2}
\end{center}
\vspace{-2em}
\end{table}

\section{Experimental Setup and Results}
\label{sec:sim_res_ch2}
The experimental setup and results for evaluating the performance of the proposed trust quantification model are delineated in this section. In general, a number of scenarios are considered to measure the accuracy of the model in observing the behaviour of SIoT objects. The term ``node'' or ``object'' in the discussion represents the SIoT objects and is used interchangeably in this section. 
\vspace{-1.5em}

\subsection{Experimental Setup}
To conduct the simulations, we used Python and utilized the SIGCOMM \cite{thlab-sigcomm2009-20120715} dataset and then conceptualized the traces of this dataset into the SIoT environment with the IoT aspect discussed in \cite{924291}. In essence, the SIGCOMM dataset contains a total of 76 objects (i.e., users) having a number of interactions with each over the span of four days, and also contains the social aspects of all the objects in terms of their friends, communities-of-interest, interactions, and other message logs. Moreover, these aspects are utilized to quantify the trust features discussed in Section~\ref{sec:proposed_model} vis-à-vis the trustor-trustee pair having at least one interaction at any given time. We have extended this dataset to incorporate 150 objects with around 20,000 interactions to efficiently realize the proposed model in terms of experimental evaluation in order to observe the long-term behaviour of the SIoT objects. The experimental analysis is carried out by separating the data into the different numbers of interactions to keep track of selected trustworthy and untrustworthy objects in order to assess the envisaged trust model and to analyze the behaviour of randomly selected objects. 

 \vspace{-1em}

\subsection{Results and Analysis}
This subsection is further divided into two parts (1) General Analysis where the trust score of randomly selected good (trustworthy) and malicious (untrustworthy) objects is analyzed with varying interaction, and (2) Behaviour Analysis where the trust-based behaviour of randomly selected nodes is discussed.  

\subsubsection{General Analysis}
\begin{figure}[t]
        \centering
        \begin{subfigure}[t]{0.32\textwidth}
              \centering
            \includegraphics[width=1.6in,height=1.25in]{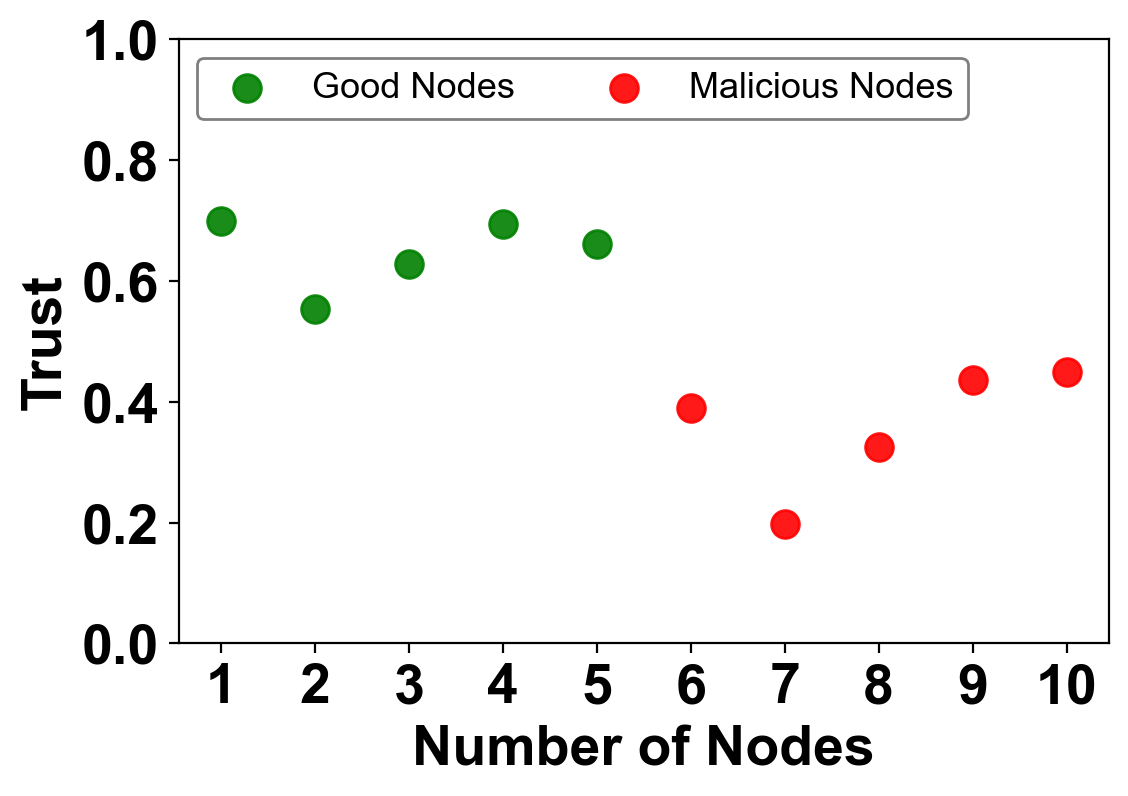}
              \caption{Trust: 4,000 interactions.}
            \label{fig:sub-first_4k}
        \end{subfigure}%
        ~
        \begin{subfigure}[t]{0.32\textwidth}
              \centering
            \includegraphics[width=1.6in,height=1.25in]{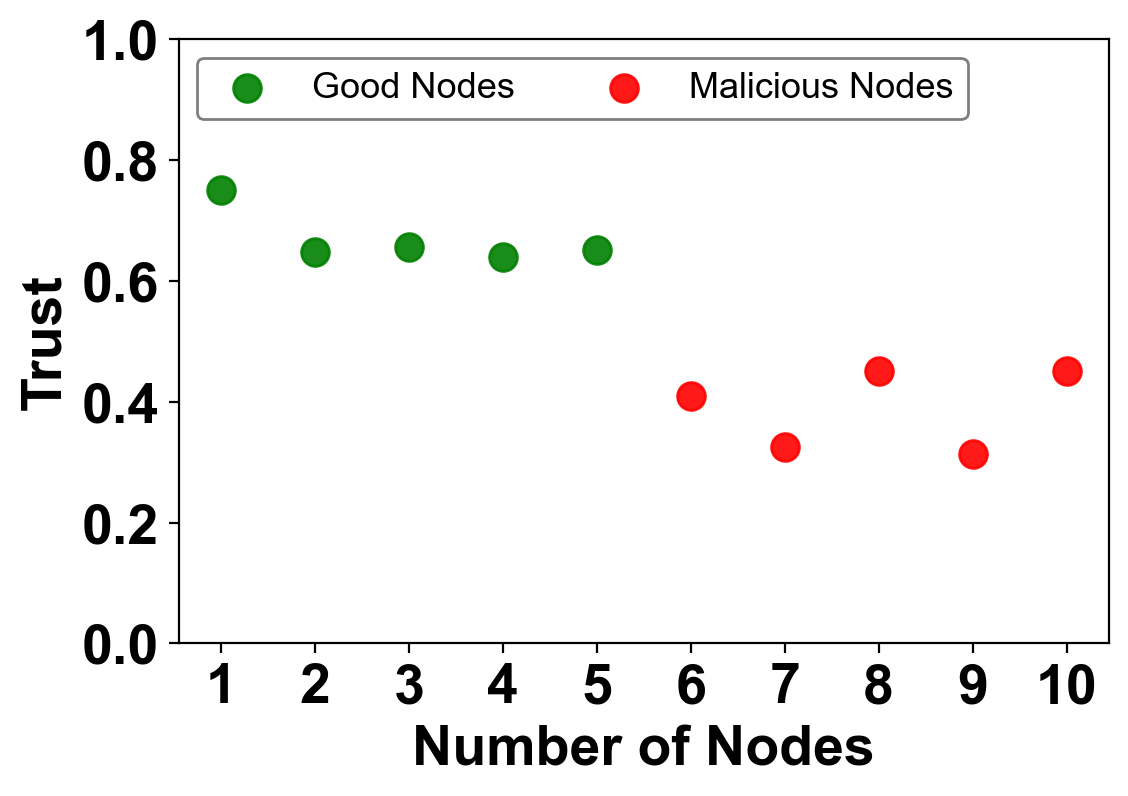}
            \caption{Trust: 8,000 interactions.}
            \label{fig:sub-first_8k}
        \end{subfigure}
        ~
        \begin{subfigure}[t]{0.32\textwidth}
              \centering
            \includegraphics[width=1.6in,height=1.25in]{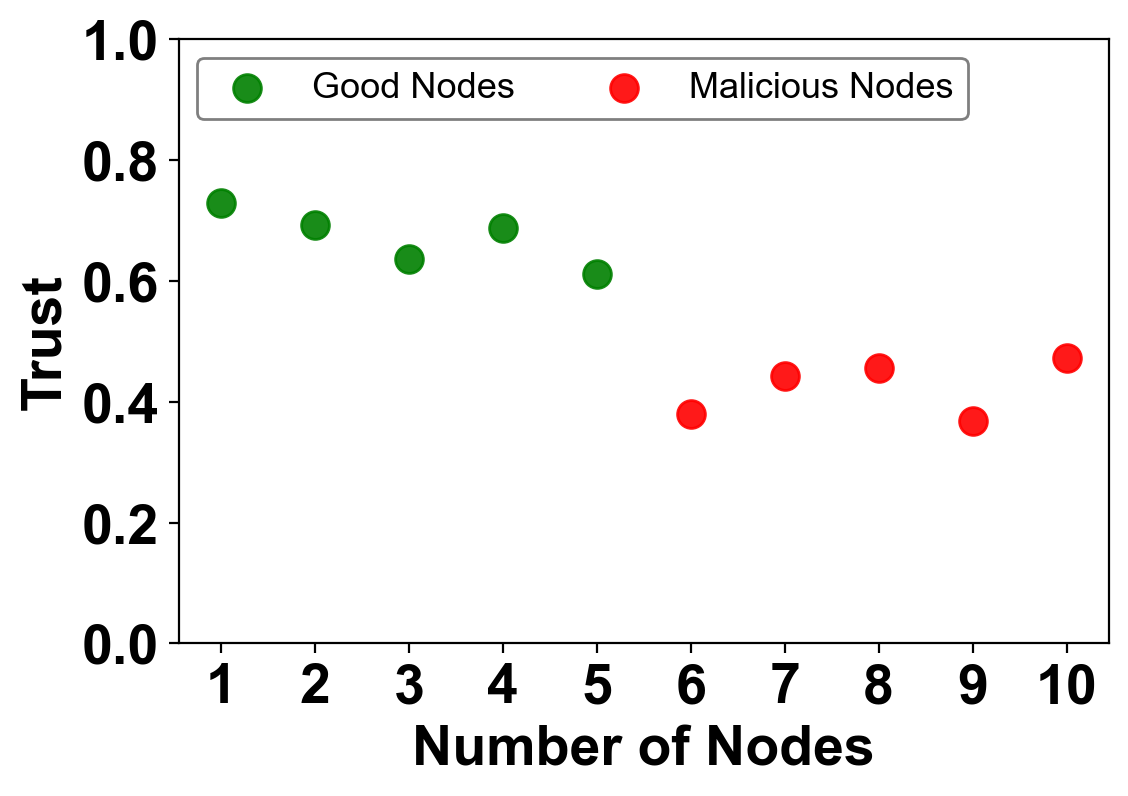}
             \caption{Trust: 12,000 interactions.}
             \label{fig:sub-first_12k}
        \end{subfigure}
        ~
        \begin{subfigure}[t]{0.33\textwidth}
              \centering
            \includegraphics[width=1.6in,height=1.25in]{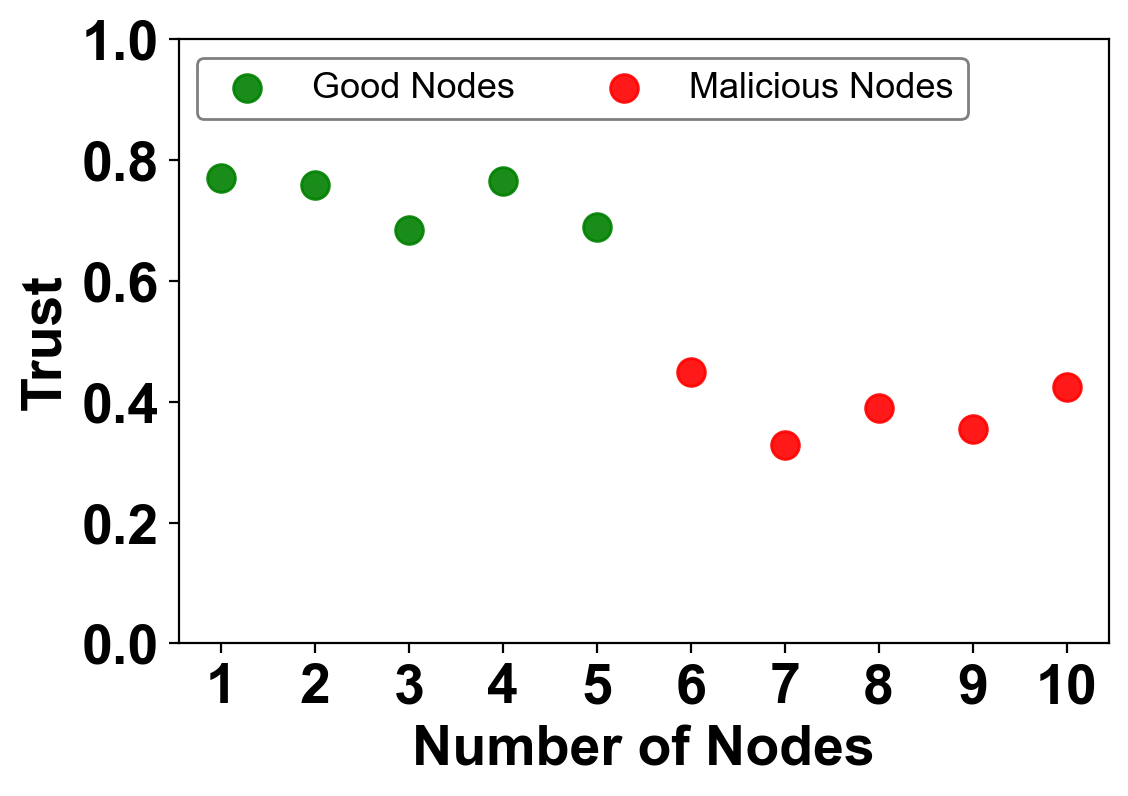}
              \caption{Trust: 16,000 interactions.}
             \label{fig:sub-first_16k}
        \end{subfigure}
        ~
        \begin{subfigure}[t]{0.33\textwidth}
              \centering
            \includegraphics[width=1.6in,height=1.25in]{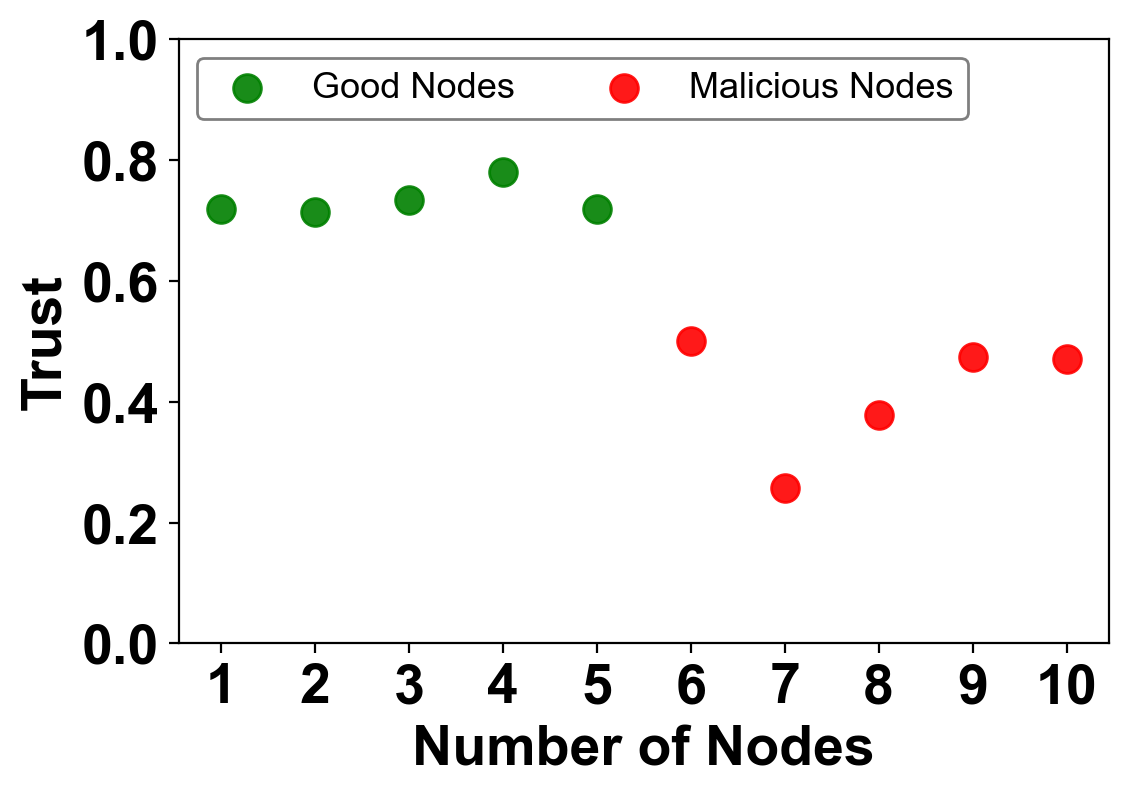}
              \caption{Trust: 20,000 interactions.}
                \label{fig:sub-first_20k}
        \end{subfigure}
        
      \caption{Trust score of randomly selected good and malicious nodes with varying interactions.}
        \label{fig:trust_score_overtime_ch2}  
        \vspace{-1em}
    \end{figure}

Figure \ref{fig:trust_score_overtime_ch2} depicts the trust score of randomly selected SIoT nodes analyzed with varying interactions. We have highlighted the behaviour of five good nodes and five malicious nodes whose trust score changes significantly with an increase in the number of interactions. It can be seen from the figure that the trust score of good nodes (trustworthy) remains above the threshold $\theta$ throughout the interactions, however, the trust score of some of these nodes remains more stable than other nodes due to the presence of malicious nodes in the SIoT network. In general, even in the presence of a number of malicious nodes, the trust score of good nodes remains above the threshold ($\theta$). Similarly, the trust score of malicious nodes remains below the threshold, nevertheless, the trust score of these nodes varies with interactions as the malicious nodes vary their behaviour to improve the trust score. It can be observed from the figures that the proposed model successfully keeps the malicious nodes below the trust threshold $\theta$. 

\begin{figure}[b]
\centering
\begin{subfigure}{.40\textwidth}
  \centering
  \includegraphics[width=\linewidth] {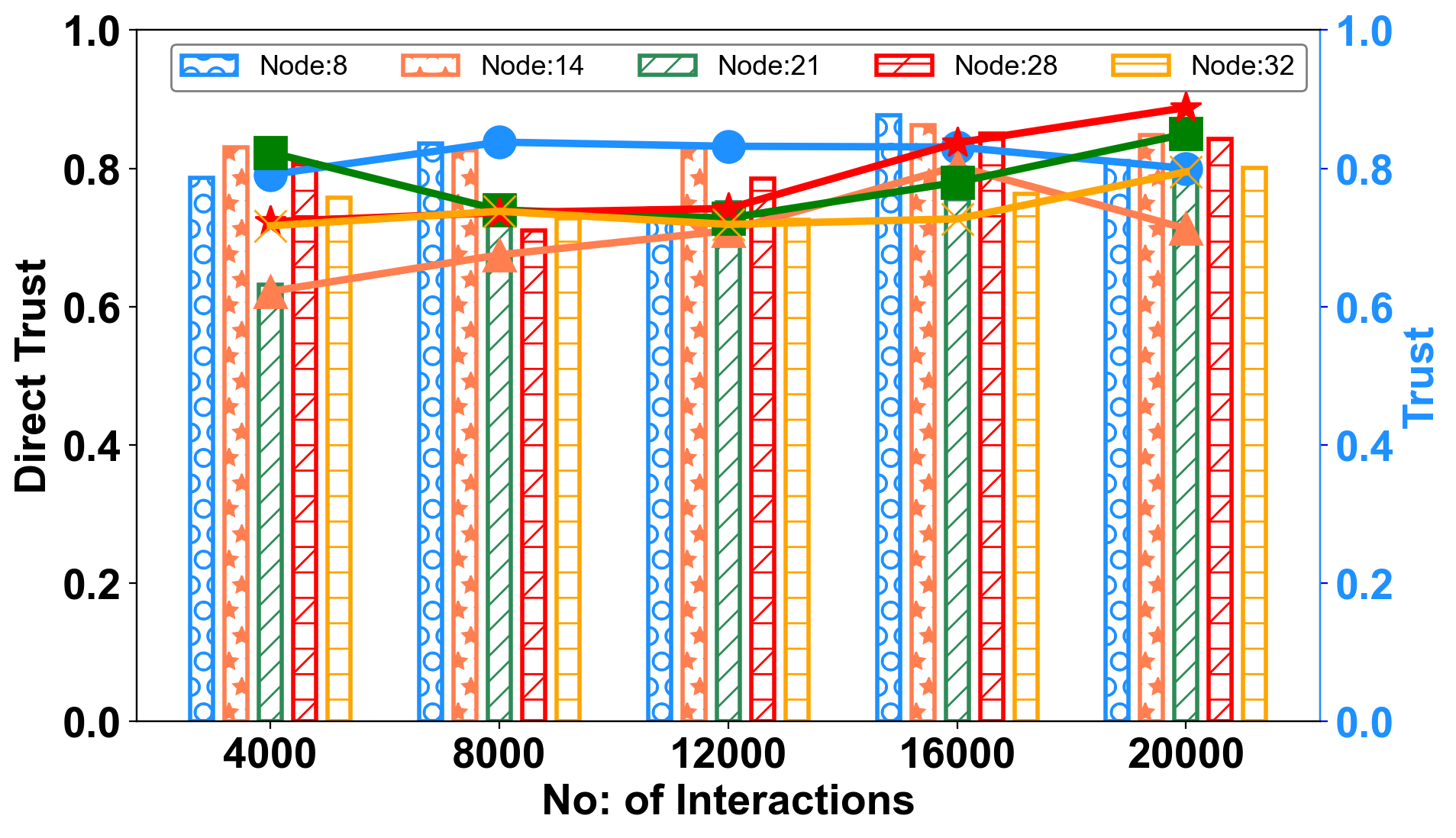}  
  \caption{Direct trust of good nodes}
  \label{fig:sub-first}
\end{subfigure}
~
\begin{subfigure}{.40\textwidth}
  \centering
  \includegraphics[width=\linewidth]{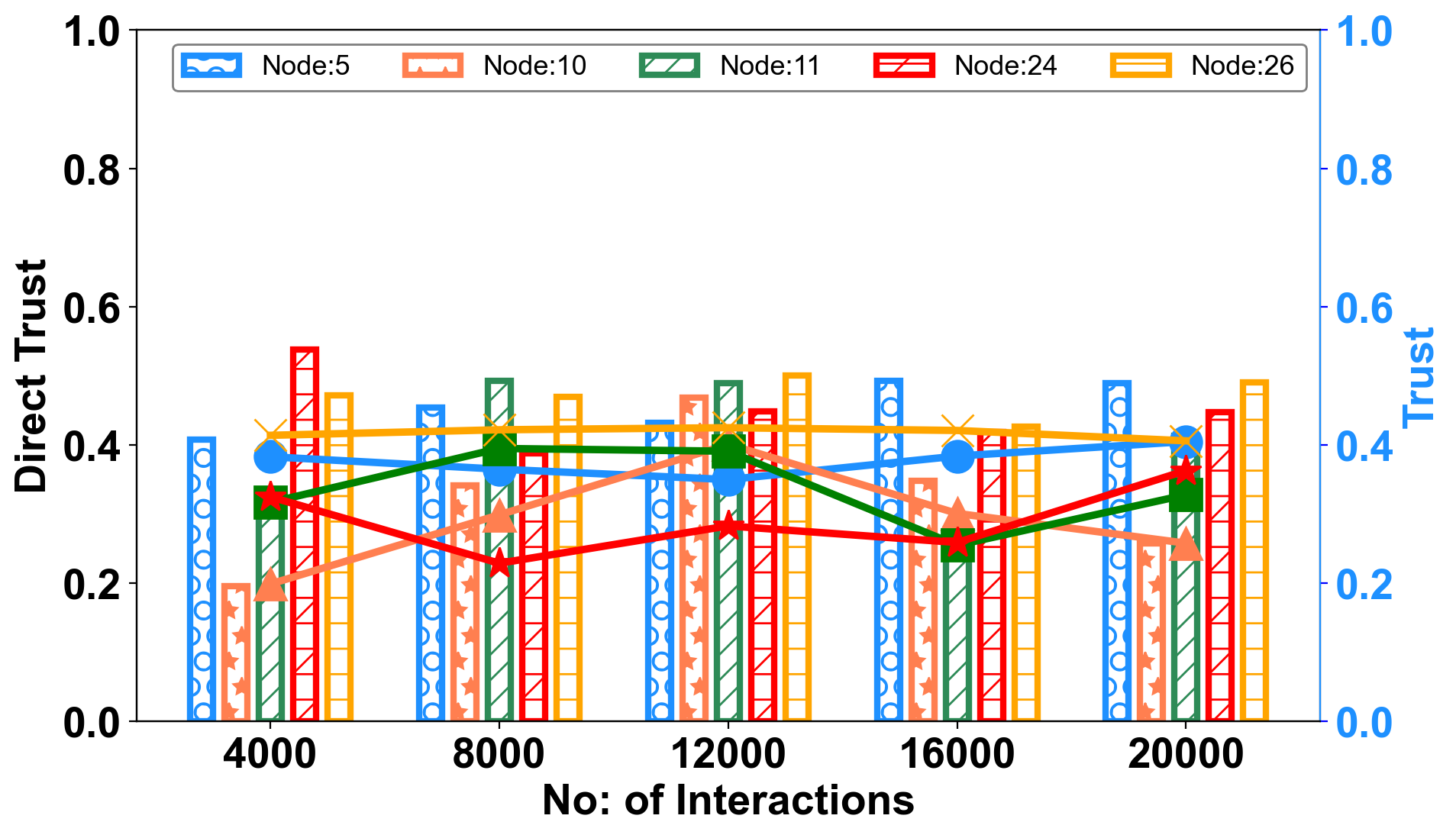}  
  \caption{Direct trust of malicious nodes}
  \label{fig:sub-third}
\end{subfigure}

\caption{Effect of direct trust on trust score of randomly selected good and malicious nodes with varying interactions.}
\label{fig:trust_features_ch2}
\end{figure}

Furthermore, Figure \ref{fig:trust_features_ch2} portrays the reason behind the variation in the trust score of the selected nodes in terms of the trust features with varying interactions. We have only considered the direct trust observation ($Trust_{DT}^t(i,j)$) to analyze the trust score of nodes as this feature is an important aspect of the proposed trust quantification model than the other trust features (i.e., $Trust_{SS}^t(i,j)$, $Trust_{R}^t(i,j)$). We observe that the trust score of good nodes (Figure \ref{fig:trust_features_ch2}(a)) remains the same from 4,000 to 12,000 as the trust features also remain intact. Nonetheless, the trust measure of these nodes increases onward due to an increase in direct trust. Likewise, the trust score of malicious nodes (Figure \ref{fig:trust_features_ch2}(b)) also depends on direct trust more than the other parameters, nevertheless, we perceive that the trust score of these nodes varies with an increase in the number of interaction and the values drop when the interactions count reaches to 20,000. As a whole, it can be observed that the trust score of both the good and malicious nodes is more inclined towards the direct trust score.

\begin{figure}[t]
    \centering
    \includegraphics[height= 5cm, width=0.6\linewidth]{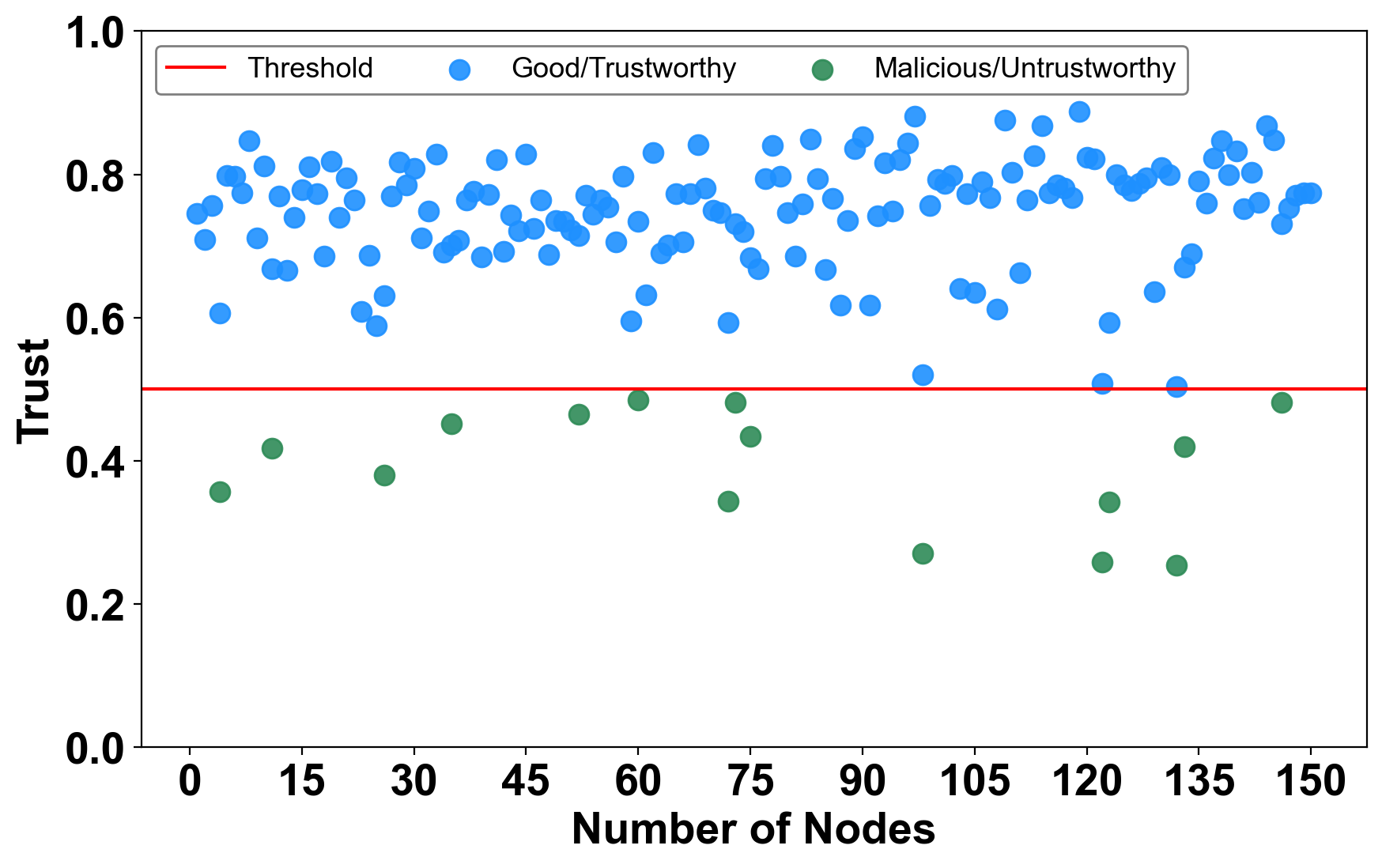}
    \caption{Trustworthiness of all the nodes in the dataset.}
    \label{fig:trust_nodes_final_ch2}
    \vspace{-1.5em}
\end{figure}

The primary objective of the proposed trust model is to efficiently quantify the trust features in order to effectively classify the objects as either trustworthy or untrustworthy. Furthermore, in order to split the object into groups, a threshold ($\theta$) is required, and these threshold values rely on a number of facets (i.e., environmental condition and application requirements).  For instance, consider a SIoT application where an object's credibility is more important than its data, as a result, there can never be a compromise on an object’s credibility in this sort of application. Hence, the threshold must be higher (i.e., $\theta>0.8$). The threshold value, however, might be lower (i.e., $theta>=0.3$) in cases when the data has a higher priority than an object's trust score. The proposed model has used the threshold value of $0.5$ (i.e., $\theta=0.5$) to categorize the objects as either trustworthy or untrustworthy in order to provide the perception that our model classifies them as such. Lastly, Figure \ref{fig:trust_nodes_final_ch2} illustrates how the proposed model classifies each node in the dataset as trustworthy or untrustworthy. Nodes with trust values more than $theta$ ($Trust \ Score > 0.5$) are designated as trustworthy, and the remaining objects are labeled as untrustworthy. The suggested model with weighted schemes (WS-1) has detected $14$ out of $15$ objects as being untrustworthy with a detection accuracy of around $94$ percent for the given figure, wherein $10\%$ ($15$ in total) of the objects are malicious or unreliable.

\subsubsection{Behaviour Analysis}

To evaluate the performance of the proposed model with a number of suggested weight schemes, in this subsection, the behaviour of randomly selected nodes in terms of trust-related attacks is analyzed with varying interactions. In particular, we have observed the following dynamic behaviours:

\begin{figure}[t]
    \centering
    \begin{subfigure}[t]{0.40\textwidth}
        \centering
        \includegraphics[width=\linewidth]{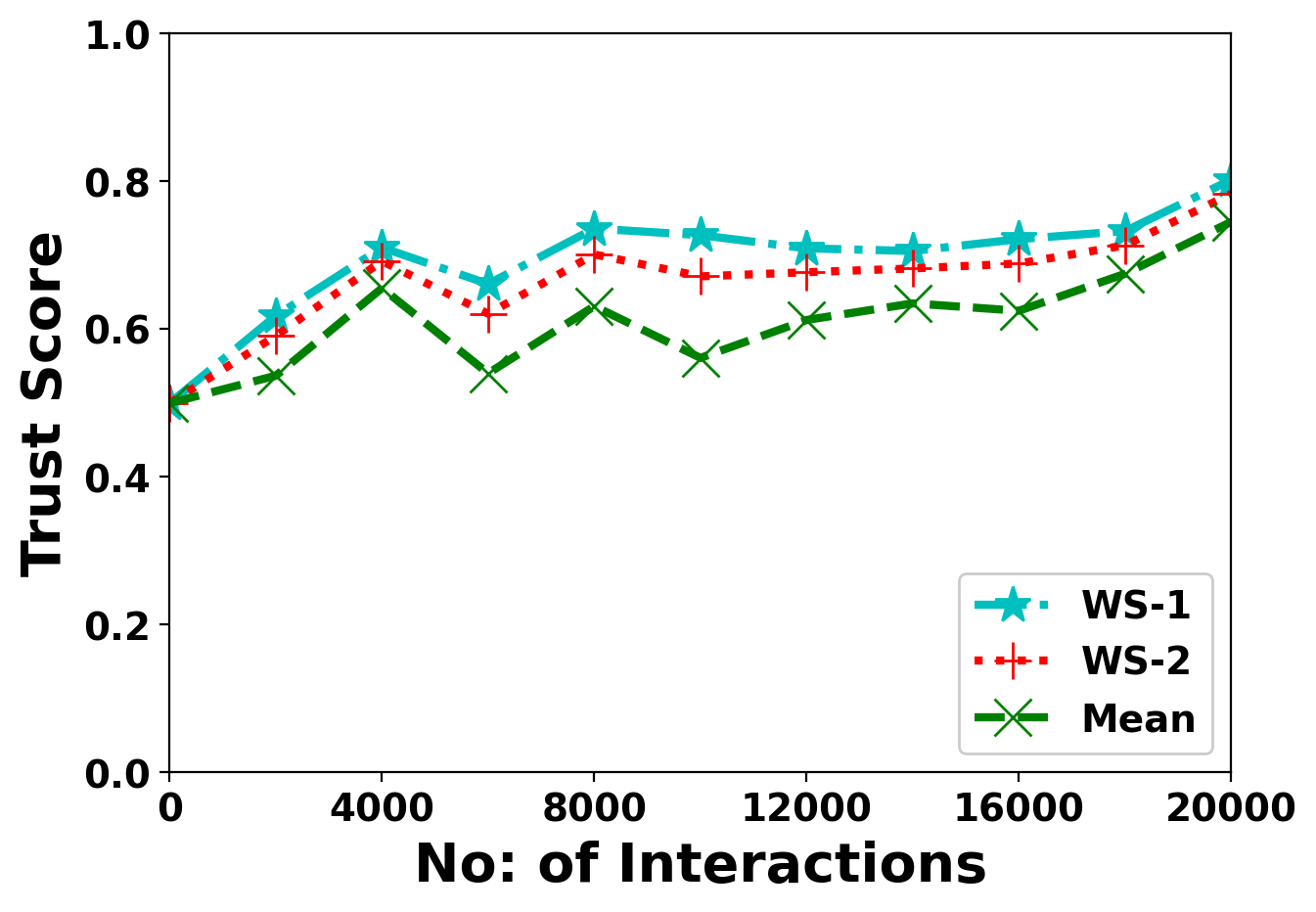}
        \caption{Trust score of a good node.}
    \end{subfigure}%
    ~ 
    \begin{subfigure}[t]{0.40\textwidth}
        \centering
        \includegraphics[width=\linewidth]{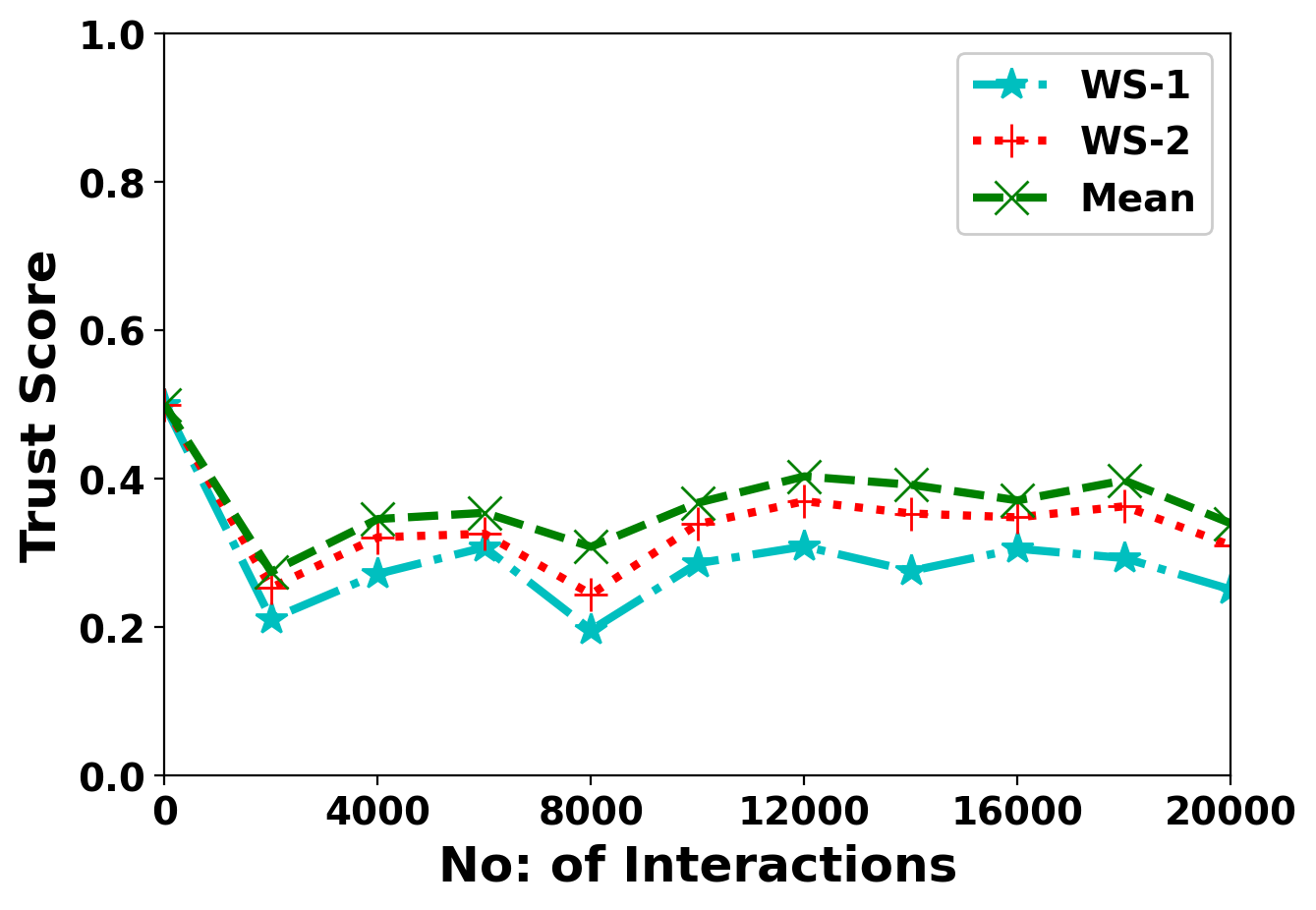}
        \caption{Trust score of a malicious node.}
    \end{subfigure}
    \caption{Trust score of a randomly selected good and malicious object with varying interactions.}
    \label{fig:beh_good_mal_ch2}
    \vspace{-1em}
\end{figure}

\begin{itemize}
   \item[1.] \textbf{Good Behaviour} -- In this type of behaviour, a node maintains its trust score throughout the interactions with a trust score above the threshold $\theta$. 
   \item[2.]\textbf{Malicious Behaviour} -- A node acts maliciously in this type of behaviour and thus, its score is always lower than the threshold.   
   \item[3.] \textbf{Good to Malicious Behaviour} -- This type of behaviour represents the change in the reputation of a node with an increase in the number of interactions, in particular, how the reputation of a node decay with interactions.   
   \item[4.] \textbf{Malicious to Good Behaviour} -- In contrast, this behaviour delineates how a malicious node develops its reputation with the increase in the number of interactions. 
   \item[5.] \textbf{On-off Behaviour} -- This type of behaviour is also known as intelligent behaviour wherein the nodes vary their reputation on and off.  
\end{itemize}

As depicted in Figure \ref{fig:beh_good_mal_ch2}(a), the proposed trust model successfully quantifies the high trust score for a good node based on its behaviour during the interaction. Similarly, Figure \ref{fig:beh_good_mal_ch2}(b) illustrates the trust score of malicious nodes, and it can be seen the trust score of this node is always low as the nodes are providing malicious services. Moreover, we have compared the trust-based behaviour with three different weight schemes and as illustrated, the WS-1 scheme outperforms the other schemes in providing a higher trust score for a good node and simultaneously providing lower trust for malicious nodes.   

\begin{figure}[t]
    \centering
    \begin{subfigure}[t]{0.32\textwidth}
        \centering
        \includegraphics[width=\linewidth]{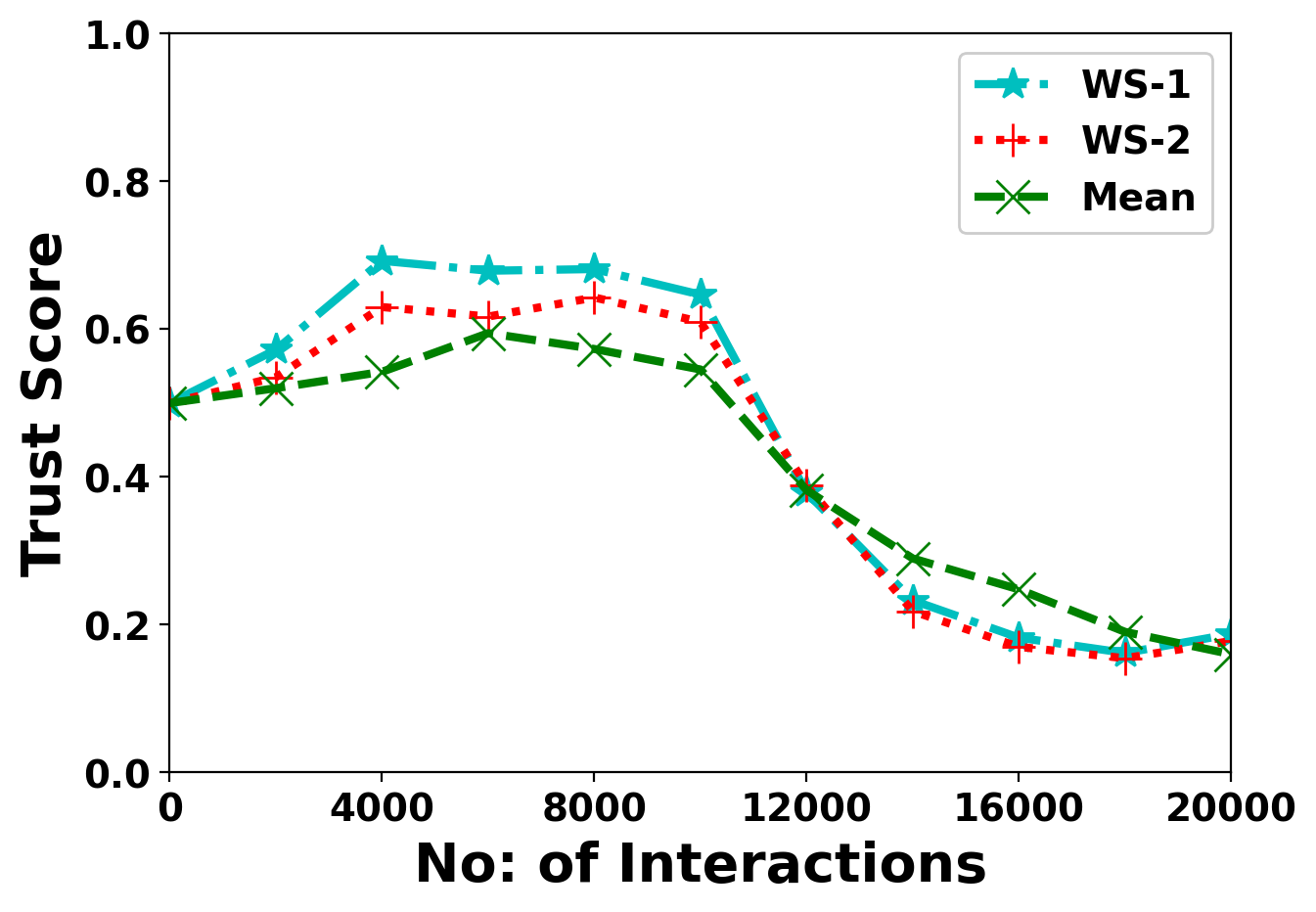}
        \caption{Good to malicious behaviour.}
    \end{subfigure}%
    ~ 
    \begin{subfigure}[t]{0.32\textwidth}
        \centering
        \includegraphics[width=\linewidth]{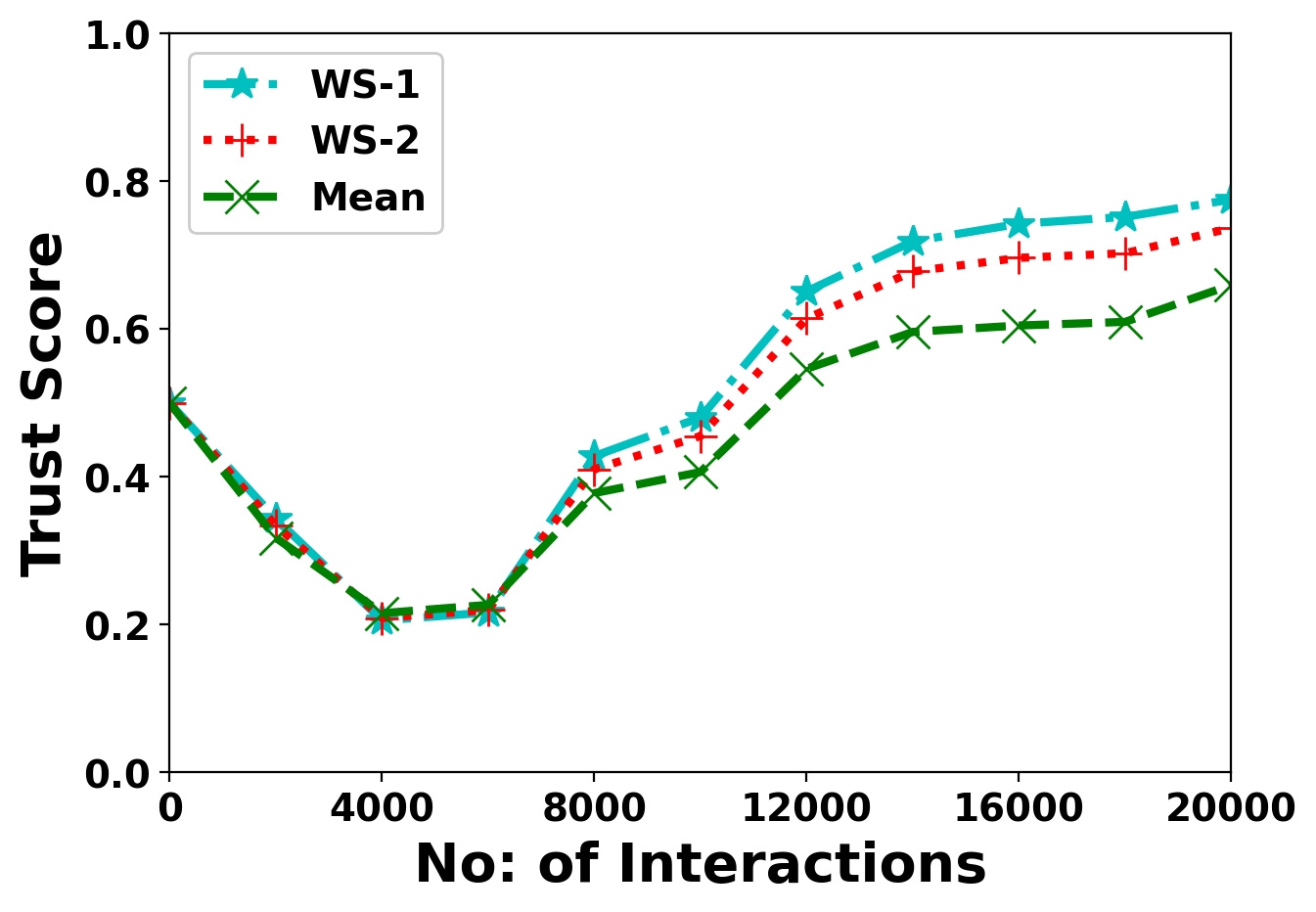}
        \caption{Malicious to good behaviour.}
    \end{subfigure}
    ~
    \begin{subfigure}[t]{0.32\textwidth}
        \centering
        \includegraphics[width=\linewidth]{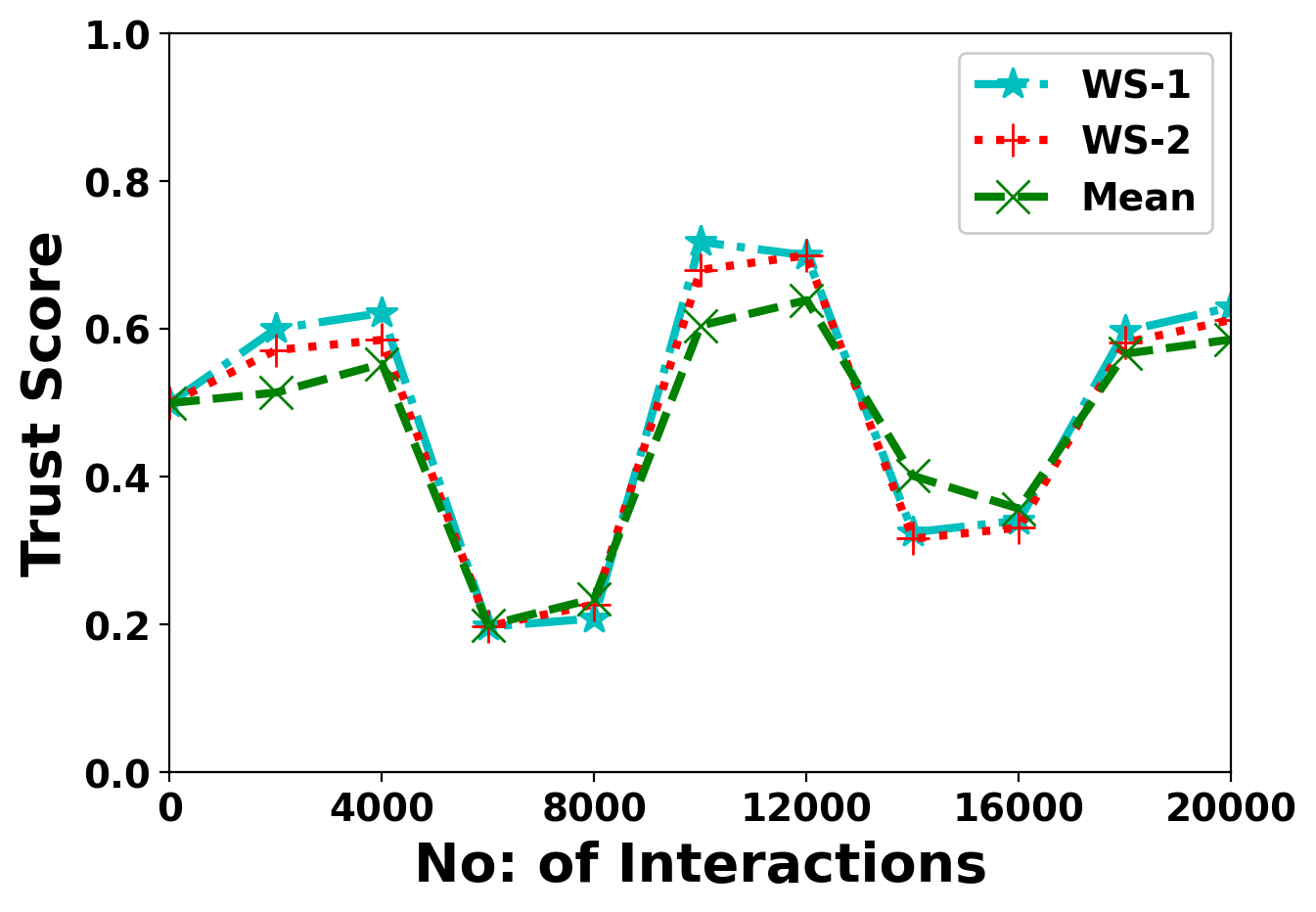}
        \caption{On-off behaviour.}
    \end{subfigure}
    \caption{Trust score of randomly selected nodes with dynamic behaviour.}
    \label{fig:dynamic_beh_ch2}
\end{figure}

\begin{figure}[t]
    \centering
    \begin{subfigure}[t]{0.40\textwidth}
        \centering
        \includegraphics[width=\linewidth]{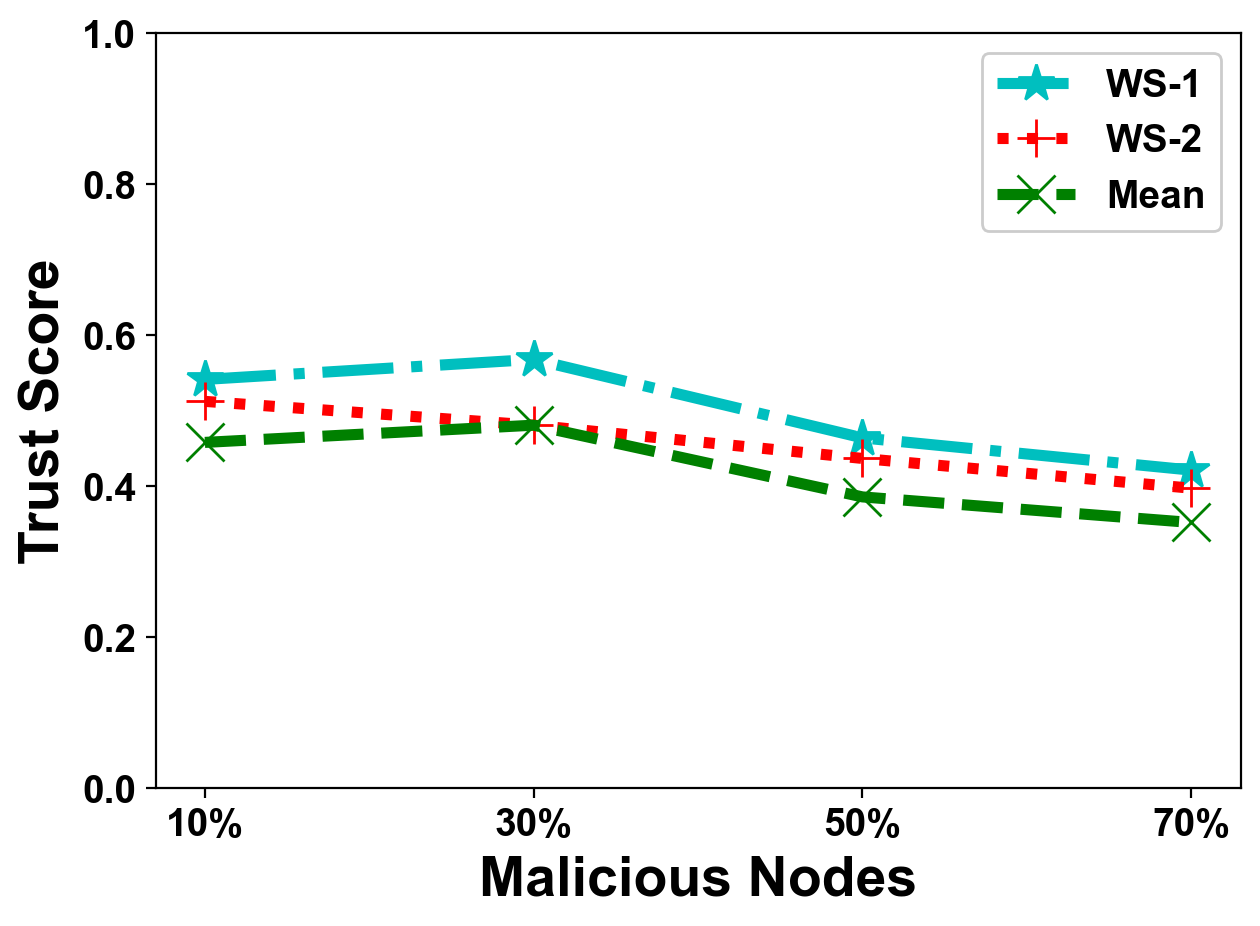}
        \caption{Trust score of a good node.}
    \end{subfigure}%
    ~ 
    \begin{subfigure}[t]{0.40\textwidth}
        \centering
        \includegraphics[width=\linewidth]{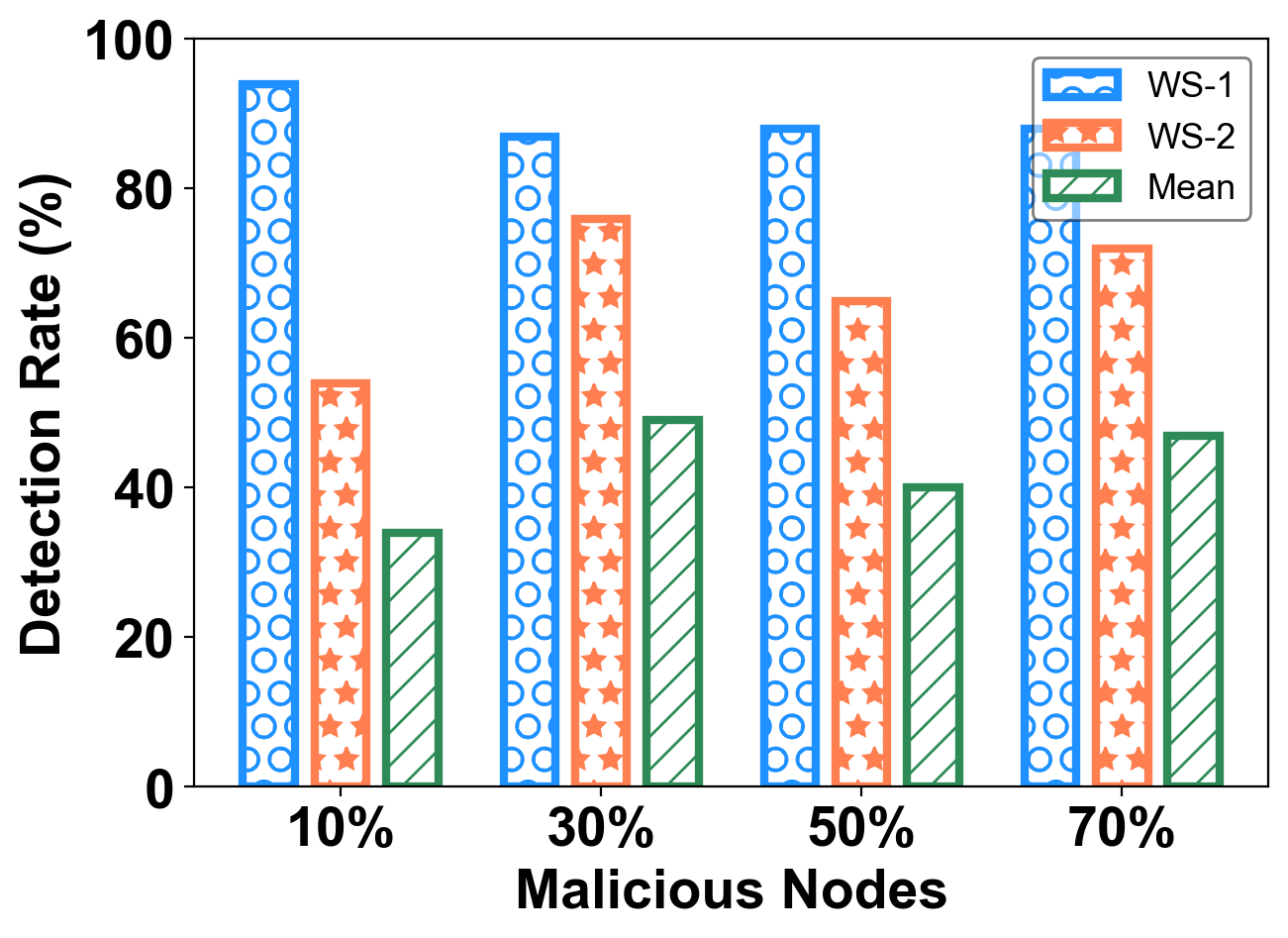}
        \caption{Trust score of a malicious node.}
    \end{subfigure}
    \caption{Trust score of randomly selected good and malicious nodes with varying percentages of malicious nodes on different weight schemes.}
    \label{fig:detecdtion_rate_ch2}
    \vspace{-1.75em}
\end{figure}

Furthermore, the dynamic behaviour of the randomly selected nodes is illustrated in Figure \ref{fig:dynamic_beh_ch2}. As can be seen in Figure \ref{fig:dynamic_beh_ch2}(a) how the behaviour (reputation) of a node varies with respect to the interactions and in particular, the reputation of the node decays and our proposed model with different weight schemes has dynamically identified the behaviour based on the interaction of the node within the network. In general, all the weight schemes have identified the behaviour successfully, however, the WS-1 outperforms the other schemes in quantifying the trust score with a higher trust score during the initial interactions and lowest trust score onwards. In contrast to Figure~\ref{fig:dynamic_beh_ch2}(a), the behaviour on how a node can enhance its reputation from malicious to a good node is portrayed in Figure~\ref{fig:dynamic_beh_ch2}(b) and it can be observed from the figure that our proposed model can identify the change in the reputation of a node with WS-1 provides the better quantification. Moreover, Figure \ref{fig:dynamic_beh_ch2} depicts the performance of the model in identifying the on-off behaviour of the objects, and the proposed model can successfully classify the good and malicious behaviour of a node with different trust scores. 

Conclusively, the aim of any trust model is to identify the untrustworthy nodes in the SIoT network in order to provide reliable services. Therefore, it is imperative to analyze the performance of the model in terms of the detection of untrustworthy nodes with a higher percentage of malicious nodes in the network and to comment on the detection accuracy of the model. Figure \ref{fig:detecdtion_rate_ch2}(a) presents the analysis of the model in terms of its success rate (trust score) with varying percentages of malicious nodes, and it can be seen that our model can converge with only a few trustworthy/good nodes in the network. Similarly, Figure \ref{fig:detecdtion_rate_ch2}(b) portrays the actual detection rate of the model and it can be seen the detection accuracy of our model is higher even in the presence of more than $50\%$ of malicious nodes. In general, the weighting scheme $WS-1$ outperforms the other schemes in detecting untrustworthy nodes.  
\vspace{-1em}

\section{Conclusion and Future Directions}
\label{sec:summ_ch2}
This paper proposes a trust quantification model that amalgamates the notion of trust in aspects of the direct trust of an object towards another object, indirect trust as a recommendation, and social similarities (community-of-interest, friendships, and the degree of co-work relationships). The trust evaluation takes place when an object, i.e., a trustor, interacts with another object, i.e., a trustee. At first, the direct trust of an object is assessed in a subjective manner by utilizing the count of positive and negative interactions. Subsequently, the trustor requests recommendations from trustworthy neighbours, and the degree of social similarity is computed as the trust feature. The final step in the trust quantification is to aggregate all the trust features, i.e., the proposed model employs a weighted sum approach for ascertaining the final trust score. Finally, the experimental evaluations demonstrate how the trust score of randomly selected trustworthy and untrustworthy objects evolve over time and how the proposed model effectively identifies the dynamic behaviour of the SIoT objects.   

In order to further investigate the precision and convergence of the proposed model in a dynamically evolving SIoT context, in the future, we intend to propose a trust model that would integrate context awareness in terms of time and environmental conditions. We further intend to employ knowledge graph embeddings to effectively amalgamate the SIoT relationships in terms of social similarities for a realistic trust assessment.

\bibliographystyle{unsrt}
\bibliography{references}

\end{document}